\documentclass[useAMS,usenatbib,usegraphicx]{mn2e}
\usepackage{rotating}
\usepackage{txfonts}

\def\lae{\mathrel{<\kern-1.0em\lower0.9ex\hbox{$\sim$}}}
\def\gae{\mathrel{>\kern-1.0em\lower0.9ex\hbox{$\sim$}}}

\title[The central stellar mass density of ETGs]
{On the central stellar mass density and the inside-out growth of 
early-type galaxies}
\author[P. Saracco et al.]{P. Saracco$^{1}$\thanks{E-mail: 
paolo.saracco@brera.inaf.it},  A. Gargiulo$^{1}$, M. Longhetti$^{1}$\\ 
$^{1}$INAF - Osservatorio Astronomico di Brera, Via Brera 28, 20121 Milano,
Italy
}
\begin{document}

\date{Accepted 2012 February 25}
\pagerange{\pageref{firstpage}--\pageref{lastpage}} \pubyear{2012}
\maketitle
\label{firstpage}
\begin{abstract}
Recent theoretical and observational studies on the assembly 
of early-type galaxies (ETGs) point towards an inside-out growth of their 
stellar mass characterized by extended low mass-density halos grown
around compact and dense cores. 
Models can form ETGs at high-z as compact spheroids that then grow in size 
through dry minor mergers.
Dry mergers would affect mainly the outskirts of the galaxy enlarging the 
size (i.e. the effective radius) keeping the inner
parts and the total stellar mass nearly unchanged. 
Hence, the central stellar mass density will not change with
time in contrast to the stellar mass density within the effective radius
which should decrease with time as the effective radius increases.
Some previous observations are interpreted as supporting inside-out 
growth as 
the central stellar mass density of high-z ETGs is found similar to that 
of local ETGs. 
In this paper we derive the central stellar mass density within a fixed
radius and the effective stellar mass density within the effective radius 
for a complete sample of 34 
ETGs morphologically selected at $0.9<z_{spec}<2$ and compare them with 
those derived for a sample of $\sim900$ local ETGs in the same mass range.
We find that the central stellar mass density of high-z ETGs spans just
an order of magnitude and is similar to that of local ETGs, as
found in previous studies.
However, we find that the effective stellar mass density of high-z ETGs
spans three orders of magnitude, exactly as the local ETGs and that it
is similar to the effective stellar mass density of local ETGs showing 
that it has not changed since $z\sim1.5$, in the last 9-10 Gyr.
Thus, the wide spread of the effective stellar mass density observed 
up to $z\sim1.5$ must originate earlier, at $z>2$. 
Furthermore, we show that the small scatter of the central mass density of ETGs
compared with the large scatter of the effective mass density is 
simply a peculiar feature of the Sersic profile and hence is independent of 
redshift and of any assembly history experienced by galaxies.
Thus, it has no connection with the possible inside-out growth of ETGs.
Finally, we show a tight correlation between the central stellar mass 
density and the total stellar mass of ETGs in the sense that the central 
mass density increases with mass as $\mathcal{M}_*^{\sim0.6}$.
This implies that the fraction of the central stellar mass of ETGs
decreases with the mass of the galaxy. 
These correlations are valid for the whole population of ETGs considered
independently of their redshift suggesting that they originate in the 
early-phases of their formation. 
\end{abstract}

\begin{keywords}
galaxies: evolution; galaxies: elliptical and lenticular, cD;
             galaxies: formation; galaxies: high redshift
\end{keywords}

\section{Introduction}
The formation and the mass assembly history of early-type galaxies (ETGs;
ellipticals and lenticulars) are central issues in the general
topic of galaxy formation, issues that are not fully understood.
Much effort has been devoted to understanding the formation and the 
evolution of ETGs in the context of hierarchical models of structure formation
(e.g. Khochfar and Burkert 2003; De Lucia et al. 2006; Hopkins et al. 2010). 
In these models the formation of ETGs is directly linked to mergers events.
The old stellar populations observed in almost all local ETGs suggest
that most of the stellar mass formed a very long time ago and not during recent
episodes of star formation (e.g. Thomas et al. 2005; Renzini 2006).
This led to conclude that mergers that formed ETGs either occurred at very 
high redshift, $z>4-5$, concurrently with or even triggering large episodes 
of star formation 
(e.g. Naab et al. 2007;  Khochfar S. and Silk J. 2006a) or 
were "dry", that is the coalescence of pre-existing old stellar 
systems (e.g. De Lucia et al. 2006).
In fact, the current and widely accepted view involves a combination of these
 two
merger types acting at different time. 
Indeed, ETGs are thought to form at high redshifts as compact spheroids
resulting from a main gas-rich merger event in which most of the stellar
mass of the galaxy forms.
They should then grow  in size mostly by the addition a low stellar mass density 
envelope through subsequent dry minor mergers at later times  
(e.g. Hopkins et al. 2009; Naab et al. 2009; Bezanson et al. 2009).
Naab et al. (2007) found that their simulated galaxies start forming
stars at $z\simeq3-5$ concurrently with the intense phase of dissipative 
merging.
In about 1.5 Gyr (at $z\sim2.5$) almost 80 per cent of their final stellar 
mass is formed in the dissipative central starburst 
(see also Sommer-Larsen and Toft 2010) justifying the old
stellar populations observed in local ETGs.
Then, the resulting compact and massive spheroid grows in size through dry 
minor mergers whose main effect would be that of re-arranging stars 
mostly those far from the galaxy center, efficiently enlarging  the size but 
keeping  the inner parts and the total stellar mass of 
the galaxy nearly unchanged (e.g. Naab et al. 2009).

This scenario was triggered by the fact that the first observed high-z 
ETGs were characterized by an effective radius smaller than the mean radius
of local counterparts with comparable mass, requiring a growth of their 
size across the time to match the size of the local ETGs.
The above scenario, often called inside-out growth, qualitatively reproduces 
this possible size evolution of ETGs discussed and debated in many studies 
(Daddi et al. 2005; Trujillo et al. 2006; Longhetti et al. 2007; 
Trujillo et al. 2007; Cimatti et al.
2008; van der wel et al. 2008; van Dokkum et al. 2008; McGrath et al. 2008;
Damjanov et al. 2009; Saracco et al. 2009; van der Wel et al. 2009; 
Trujillo et al. 2011; Damjanov et al. 2011; van de Sande et al. 2011).
However, it requires a considerable fine tuning to reproduce both the local 
scaling relations and their scatter which is significantly smaller than
the one produced by minor mergers (Nipoti et al. 2009; Nair et al. 2011).
{  In addition to this argument, there are some observations
that provide evidence that contradicts this stellar 
mass growth and the more general hierarchical scheme.
Indeed, recent studies focused on brightest cluster galaxies at
$1.2<z<1.5$ show that they 
were almost fully assembled by this time having already grown to more 
than 90 per cent of their final stellar mass at that redshift (Collins et al.
2009; Stott et al. 2010). 
In agreement with these results, studies on the evolution of the galaxy 
stellar mass function show that massive ETGs were already assembled at $z>1$ 
and that they have not grown further at lower redshift 
(e.g. Pozzetti et al. 2010). 
These results suggest that merging has had little effect on the  
stellar mass growth of ETGs in the last 9 Gyr contrary to 
hierarchical models that predict a protracted mass build-up over 
Hubble time.}
Furthermore, in the last couple of years evidence of a large number of 
ETGs at $z\sim1.5$ with sizes similar to those of local ones with 
comparable mass, here termed normal ETGs, has been accumulated  showing that most 
of the high-z early-types were not more compact than local ones
(Saracco et al. 2009; 2010; Mancini et al. 2010).

One of the differences between those works finding  almost exclusively 
compact ETGs at high-z and those finding a majority of normal ETGs is the 
definition of early-type galaxy and, consequently, the selection criterion 
used to construct the samples of ETGs, a criterion that also determines the 
completeness of the samples, possible selection bias and the redshift 
range covered.
The studies showing that the population of ETGs at high-z is composed of 
compact galaxies are based on samples of galaxies selected according
to their colors and/or on samples of passive galaxies selected according 
to their low value of the specific star formation rate, a quantity derived 
from the best fits of their spectral energy distribution with stellar 
population synthesis models.
These selection criteria allow the selectio of ETGs, according to the above 
definitions of ETG, even beyond $z\sim2.5$ (e.g. Buitrago et al. 2008; 
Cameron et al. 2011; Cassata et al. 2011).
%produce necessarely uncomplete samples of galaxies often including
The studies showing a predominance of normal ETGs at high-z are based on 
morphologically selected samples and, for this reason, are usually limited to 
a slightly lower redshift ($z\sim2$) than the former (e.g. Saracco et al. 2010;
Mancini et al. 2010).  
We will discuss the implications that these different selection criteria 
of ETGs may have on the results in later sections of this paper.
 
Concurrently with the evidence of a large number of normal ETGs at $z\sim1.5$, 
evidence of the presence  of a large fraction of compact 
ETGs in the local Universe similar to the high-z ones has emerged
(Valentinuzzi et al. 2010a; 2010b).
In fact, the number density of compact ETGs morphologically selected 
at $z\sim1.5$ is found to be consistent with the number density of ETGs selected 
in the same mass range in the local Universe (Saracco et al. 2010) showing 
that size evolution, if it exist, cannot affect the majority of the high-z ETGs,
in agreement with what is found from other analysis
(e.g. Newman et al. 2010; Stott et al. 2011).
These recent results indirectly cast some doubt on the inside-out growth
as a general mechanism of accretion of stellar mass and of growth in size.  
On the other hand, other studies focusing on the central and the effective 
stellar mass density of high-z ETGs have shown that the former is almost 
constant and similar to that observed in the local ETGs 
(e.g. Bezanson et al. 2009; van Dokkum et al. 2010; Tiret et al. 2011) 
providing one of the strongest arguments in favor of the hypothesized 
inside-out growth of ETGs throughout their life.

In this paper we focus on the central and the effective stellar mass density 
of ETGs and on their relations with other fundamental quantities.
We use both a complete sample of ETGs at $0.9<z_{spec}<2$ and a 
local sample of ETGs selected according to the same criteria as used for the 
high-z sample in order to check for possible differences.
The aim of this analysis is twofold.
First, we want to asses whether the central and the effective
stellar mass densities of ETGs are linked to the evolutionary path followed
by ETGs and hence whether they can be used to probe their assembly history or 
whether they are independent on it.
Secondly, we want to asses whether inside-out growth is a viable 
way to justify the properties observed in ETGs both at high-z and in the
local universe examining both the evolution they underwent at $z<1.5$ and
the early phases they possibly experienced at $z>2$.
In Sec. 2 we describe the data.
In Sec. 3 we define the central and the effective stellar mass densities 
probing the inside-out growth scenario at $z<1.5$. 
In Sec. 4 we use the results obtained to constrain the general 
spheroids formation scheme at very early epochs.
In Sec. 5 we define and study a scaling relation involving the central 
and the total stellar mass, scaling relation followed by ETGs independently
of their redshift.
Finally, in Sec. 6, we summarize the results and present our conclusions.
Throughout this paper we use a standard cosmology with
$H_0=70$ Km s$^{-1}$ Mpc$^{-1}$, $\Omega_m=0.3$ and $\Omega_\Lambda=0.7$.
All the magnitudes are in the Vega system, unless otherwise specified.

\section{The data set}
The sample of ETGs we used in our 
analysis is composed of 34 galaxies morphologically selected from 
the southern field of the Great 
Observatories Origins Deep Survey (GOODS-South; Giavalisco et al. 2004) 
described in Saracco et al. (2010).
{
The sample was constructed by first selecting all  galaxies brighter 
than K$\simeq$20.2 over the $\sim143$ arcmin$^2$ of the GOODS-South field and 
then by removing all the galaxies with measured spectroscopic redshift 
$z_{spec}<0.9$ and those with irregular or disk-like morphology. 
This first step of the morphological classification was undertaken through
a visual inspection of the galaxies carried out independently by two of us
on the ACS images in the F850LP band.
{  Only galaxies for which the two independent visual classifications agree 
were removed.}  
Then, on the basis of the best-fitting procedure to the observed 
profile described below,  galaxies having 
a S\'ersic index $n<2$ or clear irregular residuals resulting from the fit
were removed.
Out of the 38 early-type galaxies thus selected 34 had measured spectroscopic 
redshift $0.9<z_{spec}<1.92$ leading to $\sim90$ per cent spectroscopic 
completeness.  

We note that at least eight galaxies of our sample show weak 
 OII emission line (see Tab. 1).
The lack of X-ray emission associated to these ETGs suggests that the 
emission is most probably associated with star formation rather than with AGN.
Hence, these ETGs are not passive.
Given the S/N ($<10$) of the spectra at 
$\lambda_{rest}\simeq$3500\AA-4000\AA\ and the continuum emission
at these wavelengths,
we estimate that only emission stronger than 
$\sim10^{-18}$ erg cm$^{-2}$ s$^{-1}$ could be detected.
Thus, the star formation rate for these eight galaxies would
be in the range 0.4-2 solar mass per year, according to their redshift, 
where we used the relation 
SFR(M$_\odot$ year$^{-1}$)=$(1.4\pm0.4)\times10^{-41}$ L[OII](erg s$^{-1}$)
(Kennicutt 1998).
It is worth noting that for masses of a few times $10^{10}$ M$_\odot$ this would 
result in a specific star formation rate $-11<$log(SSFR)$<-10$ yr$^{-1}$,
higher than the SSFR often used to define passive galaxies (e.g. 
log(SSFR)$\le-11$ in Cassata et al. (2011)).
We cannot exclude that the number of non-passive ETGs in our sample is 
higher than eight.
Indeed, we note that the spectroscopic observations collected for these 
eight galaxies  (di Serego Alighieri et al. 2005; Cimatti et al. 2008;
 van der Wel et al. 2005)
 are deeper than those of the other GOODS spectroscopic 
observations.
%Moreover, we point out that the spectroscopic observations for most of the 
%sample cover the rest-frame wavelength range $lambda_{rest}<0.5$ where 
%the continuum of ETGs is extremely faint, hence noisy. 
Consequently, for many of them,in particular for 
those at $z>1.4$, spectroscopic observations do not
allow detection of OII emission lines fluxes as faint as a few
$10^{-18}$ erg cm$^{-2}$ s$^{-1}$, corresponding to 
SFR$>>1$ M$_\odot$ yr$^{-1}$ at that redshift.
Hence, the presence of weak star formation would be hardly detectable. 
We note also that ETGs with colors bluer than those often used to
select passive galaxies at $z>1$ (e.g. (R-K)$>5$, or 
(I-H)$_{AB}>2.1$, Cameron et al. 2011) are present in our sample.
All these ETGs, those non-passive and those bluer than the above color cuts, 
would not be included in a sample of ETGs selected according to their
SSFR and/or to their colors.
}
The resulting 34 ETGs of our sample are at redshift $0.9<z_{spec}<1.92$ 
and have stellar masses in the range 
$0.6\times10^{10}<\mathcal{M}_*<3\times10^{11}$ M$_\odot$
as derived by fitting their spectral energy distribution (SED) sampled by 14 
photometric points in the wavelength range 0.3-8.0 $\mu$m (Saracco et al. 2010).

\section{The central and effective stellar mass densities of ETGs}
\subsection{Surface brightness profile fitting}
{ The effective radius R$_e$ [kpc] ($r_e$ [arcsec]) of our galaxies has 
been derived by fitting a S\'ersic profile 
\begin{equation}
I(R)=I_e exp\left[-b_n\left[\left({R\over R_e}\right)^{1/n}-1\right]\right]
\end{equation}
to the observed light profile in 
extremely deep (40-100 ks) HST ACS-F850LP images.
{ The point spread function (PSF) to be convolved with the S\'ersic profile
has been constructed for each galaxy by averaging the profile of some
unsaturated stars.}
The two-dimensional fitting has been performed using \texttt{Galfit} 
software (v. 2.0.3, Peng et al. 2002).
The fitting to the galaxy profiles provided us with the axial ratio 
$b/a$ and the semi-major axis $a_e$ of the projected elliptical isophote 
containing half of the total light which is related to the circularized
(averaged) effective radius by the relation $r_e=a_e\sqrt{b/a}$. 
The effective radii R$_e$  we obtained through the two-dimensional fitting 
procedure are in the range 0.4-9 kpc.

A concern over the effective radii thus derived could arise in relation
to the different rest-frame wavelengths sampled by the F850LP filter 
given the wide redshift range of our sample.
In particular, while at $z\sim1$ the F850LP filter samples the rest-frame 
close to the B band ($\sim0.42$ $\mu$m), at $z>1.5$ this filter samples 
the rest-frame at $\sim0.3$ $\mu$m, close to the U and NUV bands.
However, the emission in the B and U bands of a galaxy is dominated by
the light coming from the same young stellar component, in contrast to the
emission at $\lambda_{rest}>0.65-0.7$ $\mu$m (R band) which is
dominated by the old stellar component.
Hence we do not expect a dependence of the size of our galaxies 
on the different rest-frame wavelength sampled, i.e. on the B-band or 
U-band according to the redshift of the galaxy.
In order to exclude the presence of this dependence we selected 
the 15 galaxies of our sample in the redshift range $0.96<z<1.1$ and we 
compared their effective radius measured in the F850LP band, sampling 
$\lambda_{rest}\simeq0.42$ $\mu$m, with the effective radius measured  
in the F606W band, sampling $\lambda_{rest}\simeq0.3$ $\mu$m.
The comparison is shown in Fig.1.
It can be seen that no dependence on wavelength is present as expected.
}
\begin{figure}
\begin{center}
%\hskip -0.2truecm
\includegraphics[width=8cm]{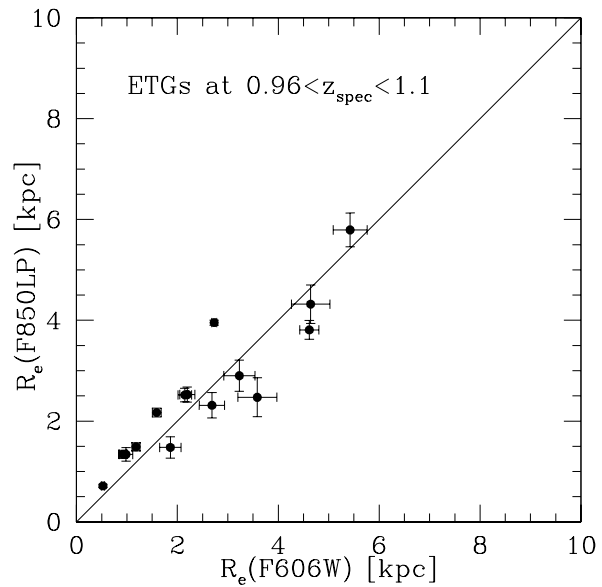} 
%\vskip -0.5truecm
\caption{The effective radius of the 15 galaxies of our sample in the 
redshift range $0.96<z<1.1$.derived in the ACS-F850LP band 
($\lambda_{rest}\sim0.42$ $\mu$m) is
compared to the effective radius derived in the ACS-F606W band
($\lambda_{rest}\sim0.3$ $\mu$m).
}
\end{center}
\end{figure}

\subsection{Stellar mass density estimates}
 In Fig. 2, the effective radius R$_e$ of the 34 ETGs of our sample  
(filled symbols) is plotted as a function of their stellar mass 
$\mathcal{M}_*$.
The solid line is the local size-mass (SM) relation of Shen et al. (2003),
while the dotted lines mark one sigma scatter.
According to the definition already used in Saracco et al. (2010),
cyan points represent the ETGs falling within one sigma from the local SM 
relation, here termed normal, while magenta points are compact ETGs, 
those falling 
at least 1 sigma below the relation.
Crosses represent $\sim900$ ETGs we selected at $z<0.05$ from the 
Wide-field Nearby Galaxy Cluster Survey (WINGS; Fasano et al. 2006; 
Valentinuzzi et al.2010a).
{
We point out that the local SM by Shen et al. is taken as reference
to define normal and compact galaxies.
It acts in the same way on any sample of ETGs considered.
For instance, we used this relation to define compact ETGs both in our
high-z sample and in the WINGS sample to compare in a consistent way
their co-moving number density (Saracco et al. 2010).}
The local sample of ETGs has been selected from the WINGS sample in the same 
stellar mass range and on the basis of their morphology, 
as done for our high-z sample. 
In particular we selected ellipticals (E) and lenticulars (S0) 
according to the morphological classification of the WINGS survey
(Fasano et al. 2012).
{
It is worth noting that WINGS effective radii have
been measured on the V-band images ($\lambda_{rest}\simeq0.5$ $\mu$m;
Pignatelli et al. 2006), very close to our $\lambda_{rest}\sim0.42$ $\mu$m).}
In the lower panel of Fig. 2, the compactness, defined as the ratio between 
the effective radius R$_e$ of the galaxy and the effective radius R$_{e,z=0}$
of an equal mass galaxy at z=0 as derived by the local SM relation, is plotted
as a function of the stellar mass for both the samples.
\begin{figure}
\begin{center}
%\hskip -0.2truecm
\includegraphics[width=8truecm]{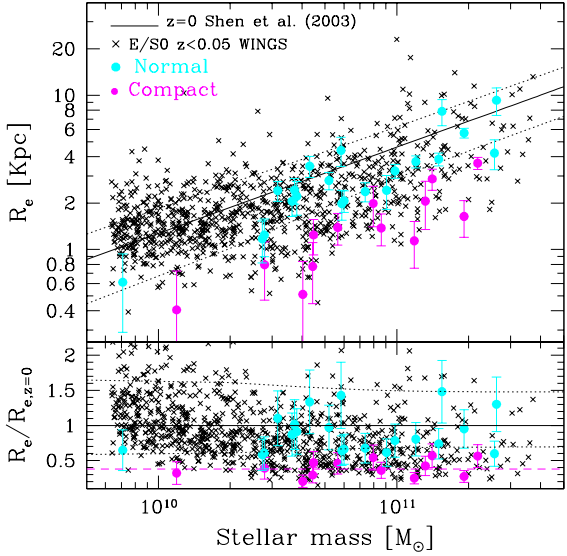} 
\caption{Upper panel: the effective radius R$_e$ of the 34 ETGs at 
$0.9<z_{spec}<2$ (filled symbols) and of the $\sim900$ ETGs selected from
the WINGS survey (crosses) is plotted as a function of their stellar mass 
$\mathcal{M}_*$.
The solid line is the local size-mass (SM) relation by Shen et al. (2003) 
with its 1 sigma scatter (dotted lines).
Cyan points are those high-z ETGs (normal) falling within one sigma from the 
local SM.
Magenta points mark high-z ETGs more compact falling at least 1 sigma below the 
relation. Lower panel: the compactness, defined as the ratio between the
effective radius R$_e$ of the galaxy and the effective radius R$_{e,z=0}$
of an equal mass galaxy at z=0 as derived by the local SM relation, is plotted
as a function of the stellar mass. Symbols are as in the upper panel.}
\end{center}
\end{figure}

Integrating the best-fitting S\'ersic profile of eq. (1) over a projected
area $A=\pi R_1^2$   we derived for each galaxy the luminosity $L_1$
interior the radius R$_1=1$ kpc
\begin{equation}
L_1=2\pi I_eR_e^2  n {e^{b_n}\over{b_n^{2n}}}\gamma(2n,x)
\end{equation}
where $n$ is the S\'ersic's index, $x=b_n(R_1/R_e)^{1/n}$ and $\gamma(2n,x)$
is the incomplete gamma function.
{ The choice of 1 kpc radius (2 kpc diameter aperture) within which to 
calculate the central stellar mass density has been made in consideration of 
the resolution of the ACS-F850LP images characterized by a FWHM$\simeq0.15$ 
arcsec corresponding to about 1.2 kpc at $z\sim1$.
{  
However, 
it is important to note that the stellar mass densities
are derived from the intrinsic galaxy profile, the S\'ersic profile.
Thus, PSF does not affect in any way the estimate of the stellar mass densities.
PSF plays a role only in the 2-dimensinal fitting to the
	observed profile as it is convolved with the Sersic profile. 
}
%Thus, a diameter aperture of 2R$_1=2$ kpc encompasses the FWHM within which
%we have no information on the light profile.
}
By replacing in eq. (2) $\gamma(2n,x)$ with the complete gamma function 
$\Gamma(2n)$ we obtained the total luminosity L$_{tot}$ 
(Ciotti 1991) and the fraction of luminosity interior to R$_1$ is then
\begin{equation}
{L_1\over L_{tot}}={\gamma(2n,x)\over \Gamma(2n)}.
\end{equation}
In our computation we assumed the analytic expression $b_n=1.9992n-0.3271$ 
(Capaccioli 1989) to approximate the value  of $b_n$.
%The value of $b_n$ used in our computation is the one given by the 
%expression $b_n=1.9992n-0.3271$ (Capaccioli 1989).
%The fraction of luminosity interior to R$_1$ is
%\begin{equation}
%{L_1\over L_{tot}}={\gamma(2n,x)\over \Gamma(2n)}
%\end{equation}
Assuming that the light profile traces the stellar mass profile and that
the mass-to-light ratio (M/L) is radially constant across the galaxy, 
%we derived
the stellar mass $\mathcal{M}_1$ 
interior to 1 kpc is given by
$\mathcal{M}_1=L_1/L_{tot}\times \mathcal{M}_*$, where $\mathcal{M}_*$
is the total stellar mass of the galaxy obtained from the SED fitting.
Because the radius R$_1$ is fixed, the stellar mass density interior 
R$_1$  is $\rho_1=c\mathcal{M}_1$, where
$c=(4/3\pi R_1^3)^{-1}=0.239$ [kpc$^-3$].
Hence, $\rho_1$ and $\mathcal{M}_1$ follows the same
behavior of the other quantities.
Under the same assumptions we also derived the effective stellar mass density
$\rho_e=0.5\mathcal{M}_*/(4/3 \pi R_e^3)$, that is the density interior 
the effective radius.
The whole set of parameters used and derived in this work for the sample of 34 ETGs
is summarized in Tab.1.
\begin{figure*}
\begin{center}
%\hskip -0.2truecm
\includegraphics[width=17cm, bb=18 400 592 718]{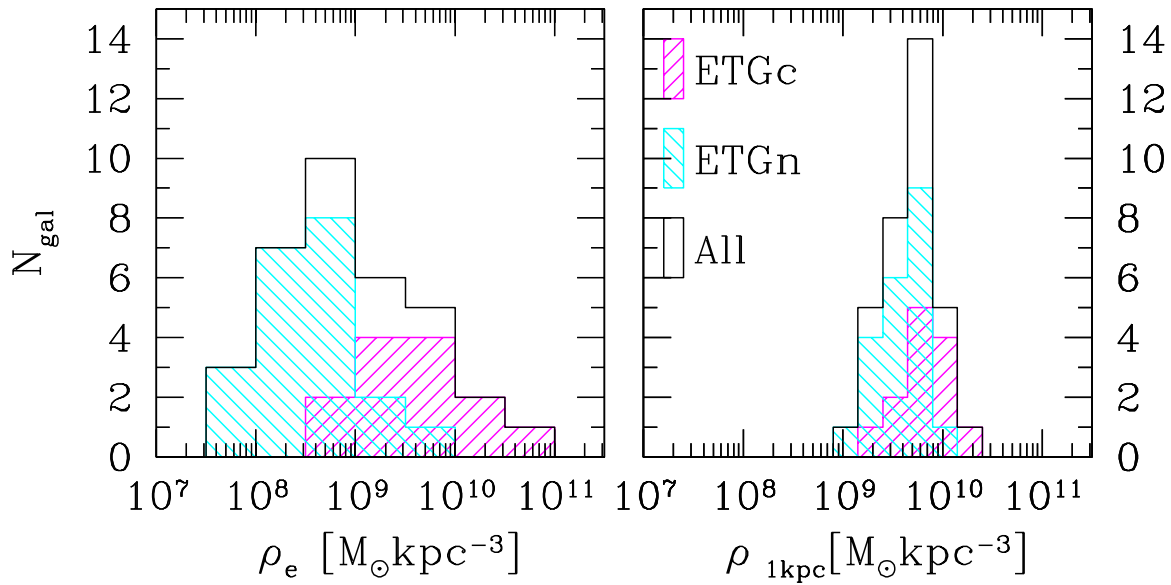} 
\vskip -0.5truecm
\caption{The distributions of the stellar mass density $\rho_e$ within the effective radius 
(left panel) and of the central stellar mass density $\rho_1$ within 1 kpc radius (right panel) 
are shown for the whole sample of 34 ETGs at $0.9<z<2$ (solid histograms). 
The shaded histograms represent the distributions of normal ETGs 
falling within one sigma from the local size-mass (SM) relation (cyan) and 
of ETGs more compact falling at least 1 sigma below the relation (magenta).}
\end{center}
\end{figure*}

{
In fact, the mass-to-light ratio could not be radially constant
for all the galaxies for the presence of color gradients 
in some high-z ETGs (e.g. Gargiulo et al. 2011, 2012; Guo et al. 2011).
We investigated whether this can affect our estimate of stellar mass
densities and our results.
(U-R)$_{rest}$ color gradients in high-z ETGs, when present, are negative 
indicating a redder color toward the center of the galaxy.
Gargiulo et al. (2012), using synthetic stellar population model fitting
to different galaxy regions, find that the main driver of such color gradients 
is the age of the stellar populations, older toward the center than in the 
outskirts with a possibly higher metallicity.
Older ages imply larger M/L values, higher metallicity lower M/L values. 
Thus the two parameters act in an opposite way regarding M/L
canceling out each other, at least partially.
Anyway, let us assume that only the age varies radially
thus producing the maximum M/L radial variation.
According to this, M/L should be larger toward the center with respect 
to the M/L we derived from the SED fitting averaged over the whole galaxy.
Consequently, both the central stellar mass density and the effective 
mass density should be slightly higher.
On the basis of the age gradients estimated by Gargiulo et al. (2012) 
we derived that the possible M/L variation is in the range 1-1.3.
Thus, the central and the effective mass densities could be larger
at most by this factor.
In particular, for those galaxies whose effective radius is comparable to 
1kpc ($\sim12$ galaxies in our sample) the effect will be exactly the same
for both the stellar mass densities, central and effective.
For the remaining galaxies whose effective radius R$_e>>1$ kpc the effect
will be slightly larger for the central stellar mass density than for the 
effective density.
Gargiulo et al. (2012) detect color gradients in 7 out of the 14 ETGs studied
and find that color gradients are not related 
to the compactness of the galaxies.
Thus, we expect that about 50 per cent of our sample show color gradients, 
independently of their compactness.
It follows that, in the hypothesis of maximum M/L radial variation (i.e.
of pure age variation)
the distributions of the central and effective stellar mass densities 
could be offset by a factor in the range 1-1.3. 
Thus, the assumption that M/L is radially constant for all the galaxies
does not introduce any significant systematic in our analysis even
in the case that it is not strictly true.
}

In Fig. 3 the distributions of the stellar mass density $\rho_e$ within the
effective radius (left panel) and of $\rho_1$ within 1kpc radius
(right panel) are shown.
The black solid-line histogram represents the distribution of the whole 
sample of 34 
ETGs while the shaded histograms show the distribution of normal ETGs (cyan) 
and of compact ETGs (magenta).
The distribution of the effective stellar mass density $\rho_e$ of 
Fig. 3 (left panel) varies by about three orders of magnitude, 
from $\sim10^8$ M$_\odot$ 
kpc$^{-3}$ to $\sim10^{11}$ M$_\odot$ kpc$^{-3}$ with a median value 
$\bar{\rho_e}=7\times10^8$ M$_\odot$ kpc$^{-3}$, reflecting the large
variation (a factor $\sim$8) of R$_e$ for ETGs of the same stellar mass
visible in Fig. 2.  
In contrast, the distribution of the central mass density $\rho_1$ shown
in the right panel of Fig. 3  spans just an order of 
magnitude centered at 
$\bar{\rho_1}\simeq 5\times10^9$ M$_\odot$ kpc$^{-3}$.
The values of $\rho_1$ are actually within a factor 2 
from the median value as  noted 
by other authors (e.g. Bezanson et al. 2009; Tiret et al. 2011).
Analogously, while the median value 
$\rho_e$ of normal ETGs ($3.5\times10^8$ M$_\odot$ kpc$^{-3}$) differs by 
a factor 10 from the median value $\rho_e$ of compact ETGs 
($3.9\times10^9$ M$_\odot$ kpc$^{-3}$) in the case of $\rho_1$ the median 
values differ by less than a factor 2 
($4\times10^9$ $vs$ $7\times10^9$ M$_\odot$ kpc$^{-3}$).
\begin{figure*}
\begin{center}
%\hskip -0.2truecm
\includegraphics[width=12.5cm]{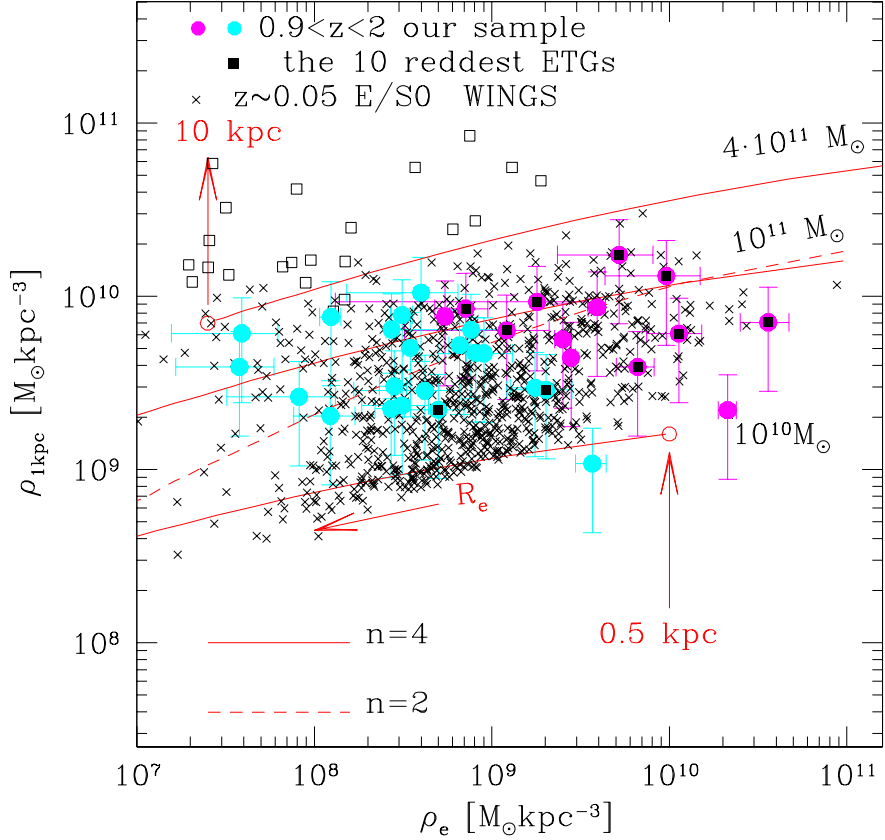} 
\vskip -0.5truecm
\caption{The stellar mass density within 1 kpc radius $\rho_1$ is plotted
versus the effective stellar mass density $\rho_e$.
Large filled dots represent the ETGs of our sample at $0.9<z<2$ (cyan normal ETGs,
magenta compact ETGs).
{ Black filled squares superimposed to filled dots mark the 10 reddest
galaxies of our sample, i.e. galaxies with color (F850LP-K)$_{AB}>2.4$.} 
Black crosses represent the local ETGs selected from the WINGS sample.
The open squares represent the 23 super-massive ($0.3-5\times10^{12}$ M$_\odot$)
ETGs selected by Bernardi et al. (2006; 2008) and studied by Tiret et al.
(2011). 
The red lines represent the locus occupied on the [$\rho_e,\rho_1$] 
plane by galaxies described by a Sersic profile with index $n=4$, 
effective radii in the range $0.5<R_e<10$ kpc and stellar masses 
10$^{10}$ M$_\odot$ (lower line), $10^{11}$ M$_\odot$ (middle line) and 
$4\times10^{11}$ M$_\odot$ (upper line) respectively.
The dashed line is obtained for an index $n=2$ and a mass
10$^{11}$ M$_\odot$.
}
\end{center}
\end{figure*}
This information is summarized in  Fig. 4 where the central
mass density $\rho_1$ of the 34 ETGs of our sample is plotted versus the 
effective mass density $\rho_e$ (filled circles).
Our data are best-fitted by the relation  
$\rho_1\propto\rho_e^{0.1\pm0.1}$, in agreement with Tiret et al. (2011) 
(to an increase by two orders of magnitude of $\rho_e$ corresponds an 
increase by a factor 1.5 of $\rho_1$) even if
the correlation between the two quantities is not 
statistically significant being the probability of the Spearmen rank 
test $P=0.27$.
The small variation of the central mass density of ETGs with respect 
to the large variation of the effective stellar mass density is often 
used to advocate the inside-out growth of stellar mass experienced by ETGs
through time (e. g. Bezanson et al. 2009; van Dokkum et al. 2010).
In this scenario ETGs, after having assembled at high-z as compact/dense 
spheroids, gain a low stellar mass density envelope through subsequent  
dry minor mergers at later times growing efficiently in size and less in mass 
and reaching at $z=0$ the typical size of local ETGs
(e.g. Hopkins et al. 2009; Naab et al. 2009).
In a plot like the one shown in Fig. 4, high-z ETGs would populate the 
right-hand side  (high effective stellar mass densities) of the 
[$\rho_e,\rho_1$] plane 
and should migrate towards lower effective stellar mass density, the 
left-hand side, at lower redshift keeping almost constant the central 
density $\rho_1$.
%The fact that the mass density within a fixed radius, e.g. within 1 kpc, 
%of high-z
%ETGs is similar to the one of local ETGs is commonly thought to support 
%the inside-out scenario.
In fact, some works based on incomplete sample of high-z ETGs pre-selected 
on the basis of their red color show this kind of segregation
(e.g. Bezanson et al. 2009).
On the other hand, it has been shown that this selection criterion 
is strongly biased toward compact ETGs (see Fig. 2 in Saracco et al. 2010).
{ To directly verify how color selection affects this plot, we selected 
the 10 reddest galaxies of our sample corresponding to a color 
(F850LP-K)$_{AB}>2.4$.
These galaxies are marked by a black filled square in Fig. 4.
It is evident that they preferentially populate the right-hand side
of the plot and thus that a color selection produce the segregation
above.
}
In contrast, it can be seen that the 34 ETGs morphologically selected 
at $z\sim1.5$ of our sample are not segregated in the 
right-hand side of the plot but 
rather they span three order of magnitude in effective stellar mass density 
in spite of their high redshift. 
{ 
Furthermore, in Fig. 5 the effective radius, the stellar mass, the central and 
effective stellar mass density of the 34 galaxies of our sample are plotted as a 
function of their redshift. 
It can be seen that there are no significant correlation between these 
quantities and the redshift, as also confirmed by the Spearmen rank test.
Hence, our analysis and our conclusions cannot be affected by any possible 
dependence on redshift of these quantities.
Actually, Fig. 5 suggests also that there is no significant evolution
of R$_e$ and $\rho_e$ over the redshift range $0.9<z<2$.
}
Thus, because at $z\sim1.5$ co-exist compact ETGs and normal ETGs, if we 
accept the framework of the inside-out accretion of stellar mass we must 
conclude that the latter must have increased their radius at earlier epochs,
at $z>2$,
while the former will increase their size later, at lower redshift.
In the next section we compare this observational evidence with the 
inside-out growth accretion proposed to assemble ETGs.

\section{Probing the inside-out growth of ETGs}
\subsection{Are the central and the effective stellar mass densities 
related to the assembly history of ETGs ?}
Whether or not a fraction of the local ETGs have undergone a similar 
mass accretion in the last 9-10 Gyr ($z<1.5-2$), to 
properly consider the observed variations of the central and the effective
stellar mass densities of ETGs, one must take into account the following
property:
if the light profile of a galaxy and
hence its stellar mass profile is represented by a Sersic profile 
(eq. 1) with index $n>2$ then the  mass density within a fixed radius 
(e.g. 1 kpc) and the effective stellar mass density are related 
 exactly as shown by the data in Fig. 4, 
independently of redshift and of mass accretion experienced.
This is shown in Fig. 4 by the red solid lines.
The lower solid line represents the locus occupied on the [$\rho_e,\rho_1$] 
plane by galaxies described by a Sersic profile with index $n=4$, 
having stellar mass 10$^{10}$ M$_\odot$ and effective radii in 
the range $0.5<R_e<10$ kpc.
The middle and the upper solid lines represent galaxies with the same 
parameters but having a stellar mass $10^{11}$ M$_\odot$ and 
4$\times10^{11}$ M$_\odot$ respectively. 
We see also that the lower the Sersic index the stronger the 
correlation between $\rho_1$ and $\rho_e$ as shown by the steeper dashed
line obtained for an index $n=2$ (normalized to  $10^{11}$ M$_\odot$).
%Indeed, the dashed red line obtained for an index $n=2$ and a stellar mass 
%$\times10^{11}$ M$_\odot$ shows a steeper relation then those obtained
%with $n=4$ (solid lines)
%the range of mass covered by our sample.
%The density $\rho_1$ is plotted versus the effective stellar mass
%density $\rho_e$ calculated for decreasing values of the effective radius in 
%the range $0.5<R_e<10$ kpc.
Both the stellar mass range and the effective radius range considered to
compute the curves are those covered by our data.
We see that the effective stellar mass density of these simulated galaxies
spans about three orders of magnitude while the 
central mass density $\rho_1$ is constrained within an order of magnitude.
Thus, any galaxy described by a set of parameters [$n$, $\mathcal{M}_*$, R$_e$]
within the above intervals ($n>2$, $\mathcal{M}_*=0.1-4\times10^{11}$ M$_\odot$,
R$_e=0.5-10$ kpc) will lie between the upper and the lower lines 
independently of redshift and of accretion history.
In fact, our high-z ETGs fall between the two lines and spans the range 
bound by the two curves.
In order to verify the independence of redshift of this evidence we used 
the $\sim$900 early-type galaxies at $z<0.05$ selected from WINGS survey.
We computed the central mass density  $\rho_1$ and the 
effective mass density $\rho_e$ of these ETGs using the same procedure
described above and used for our high-z sample.
The $\sim$900 local ETGs are shown in Fig. 4 as crosses.
As expected, they fall in the region bounded by the two red lines 
covering the same distribution as the high-z sample.
%Also, we see that the higher the Sersic index the weaker the 
%correlation between $\rho_1$ and $\rho_e$. 
%Indeed, the dashed red line obtained for an index $n=2$ and a stellar mass 
%$\times10^{11}$ M$_\odot$ shows a steeper relation then those obtained
%with $n=4$ (solid lines). 
Fig. 4 also shows the 22 super-massive ($0.3-5\times10^{12}$
M$_\odot$) ellipticals  
selected by Bernardi et al. (2006; 2008) on the basis of their high 
velocity dispersion ($\sigma_v>330$ km/s) and studied by Tiret et al. (2011)
(open squares).
It can be seen that these ETGs follow the same behavior shown by the
other data: a central stellar mass density almost constant (but centered
at a higher value according to the higher stellar mass range) 
and a large variation of $\rho_e$.
Finally, it is also important to note from Fig. 4 that the effective 
stellar mass density, the quantity which is expected to change dramatically 
(as the cube of the variation of R$_e$)
in the inside-out scenario from high-z to low-z, is not different in the 
two samples considered, in agreement with what is shown by our high-z 
sample in the lower panel of Fig. 5. 
This is more clearly shown in Fig. 6 where the distribution of the 
effective stellar mass density of our sample at $0.9<z<2$ is compared to 
the distribution of $\rho_e$ of the WINGS ETGs. 
The two distributions are equal as confirmed by the Kolmogorov-Smirnov test 
which provides a probability P=0.85.
Thus, these results show two main facts. 
The first one is that, all the ETGs independently of redshift and of their 
accretion history have the mass density interior to a fixed radius within a 
narrow range of values if compared to the stellar mass density within the
effective radius.
{ This fact reflects a peculiar feature of the true intrinsic profile 
shape that they follow which is very well reproduced by a Sersic profile  
(see also Graham and Driver 2005).
Since this behavior depends on the intrinsic profile shape of ETGs, it 
follows that also non-parametric estimators, e.g. Petrosian
radius, would return qualitatively the same results.
}
Thus, whether or not an ETG formed as a compact spheroid its central
stellar mass density will fall within a narrow range of values
depending only on its total stellar mass.
It does not depend on any past assembly history and has no connection with the 
possible inside-out growth.
The other important evidence is that the effective stellar mass density
of early type galaxies, the quantity expected to decrease dramatically in
the inside-out growth scenario as the effective radius increases, has not 
changed in the last 9-10 Gyr.
This implies that the large spread of $\rho_e$ and of R$_e$ observed both 
in the local universe and up to $z\sim1.5-2$, as shown in figures 2, 3 and 4,
must originate earlier, at $z>2$.
\begin{figure}
\begin{center}
%\hskip -0.2truecm
\includegraphics[width=8cm]{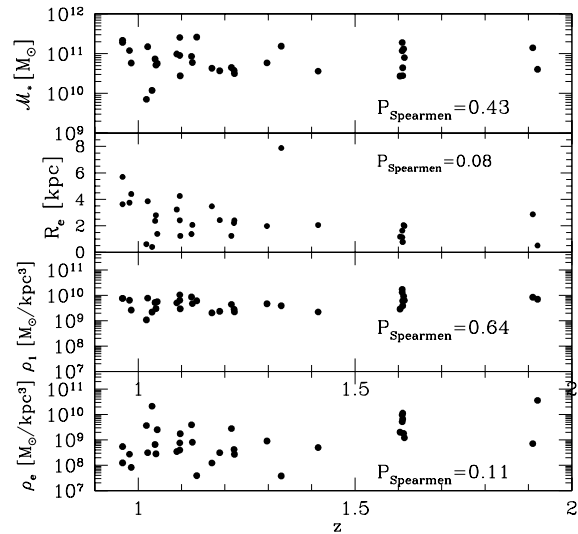} 
\caption{From top: the stellar mass $\mathcal{M}_*$, the 
effective radius R$_e$, the central stellar mass density $\rho_1$ and the 
effective stellar mass density  $\rho_e$ of the 34 ETGs of our sample 
at $0.9<z_{spec}<2$ are plotted as a function of their redshift.
No significant correlation is present. 
P$_{Spearmen}$ is the probability that the two quantities 
are not correlated as resulting from the Spearmen rank test.
}
\end{center}
\end{figure}

\begin{figure}
\begin{center}
%\hskip -0.2truecm
\includegraphics[width=8.5cm]{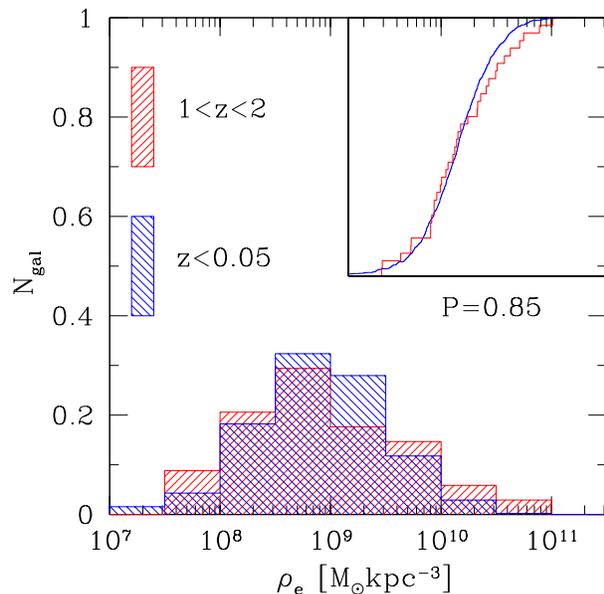} 
\vskip -0.5truecm
\caption{The distribution of the effective stellar mass density of
ETGs at $0.9<z<2$ (red histogram) is compared to the distribution
of the effective stellar mass density of the local ETGs selected from 
the WINGS survey (blue histogram). 
The effective stellar mass density is the quantity expected to dramatically 
change in the inside-out growth scenario of ETGs.
The two distributions are equal as confirmed by the Kolmogorov-Smirnov test.}
\end{center}
\end{figure}

\subsection{Stellar mass densities and the evolution of ETGs at $z>2$: 
toward the early-phases of their assembly}
Figures 4 and 6 clearly show that the properties of the local ETGs in
terms of stellar mass and stellar mass densities were already in place 
at $z\sim1.5-2$.
For this reason such properties must necessarily be the product
of the evolutionary path followed by ETGs at $z>2$, in the previous 2.5-3 Gyr
i.e. during the assembly of most of their mass.
Before making any hypothesis, let's see whether the general scheme of 
compact spheroids formation and evolution described above can justify the 
spread of the effective stellar mass density observed at $z\sim1.5-2$.
From figures 2 and 4 we see that, at a given mass, there are galaxies with 
effective radii (effective stellar mass densities) differing even by a 
factor 5-6 (more than hundred) even if all of them have nearly the 
same central mass density.
Thus, in the hypothesis that ETGs formed first as compact spheroids,
those galaxies that at $1<z<2$ have large radii (5-6 times larger) 
with respect to others with equal mass, had to increase their size at 
an earlier epoch, at $z>2$.
Hence, the inside-out accretion should have been extremely efficient at $z>2$,
in the first 2-3 Gyr of the universe to produce so many large galaxies in
so short time.
Some authors have already discussed this point in 
different contexts (e.g. Nipoti et al. 2009; Newman et al. 2011; 
Saracco et al. 2011) concluding that the required dry merging  does not 
seem to be consistent with model predictions.
Here, we can directly give an estimate of the merger rate needed on the
basis of the observed data.
Let us consider a galaxy of our sample having effective radius
(effective stellar mass density) 4 ($\sim$60) times larger than another galaxy
of equal mass at the same redshift.
Let us assume that dry minor merger is very efficient in increasing the size
of galaxies, as efficient as proposed by Naab et al. (2009) 
where the effective radius grows by the square of the change in mass 
instead of linearly.
The larger galaxy has already increased its radius
by a factor 4 during its life at $z>2$.
To increase the radius by a factor 4 
the larger galaxy had to double at least its mass at $z>2$.
In the case of mergers involving galaxies with masses, e.g.,  
$M:m\le1:5$ (1:10), the galaxy had to undergo five (ten) mergers 
in the previous $\sim2.5$ Gyr (between $2<z<6$) to double its mass.
Thus, the resulting merger rate per galaxy would be about 2 (4) mergers per 
Gyr.
If the final stellar mass at $z\sim1.5-2$ of the larger galaxy was 
$10^{11}$ M$_\odot$
the progenitor should have a mass $M\sim5\times10^{10}$ M$_\odot$ at $z>2$.
Using the merger rate calculator by Hopkins et al. (2010)
we estimated at $z=2.5$ a merger rate per galaxy of about
$6\times10^{-3}$ Gyr$^{-1}$ for masses in the range 
5$\times10^{10}-10^{11}$ M$_\odot$.
We considered dry merging, i.e. mergers involving gas fractions up to  
10\% of the total mass (gas+stars).
The predicted merger rate would be even lower for lower masses 
($6\times10^{-4}$ Gyr$^{-1}$ for masses $<5\times10^{10}$ M$_\odot$), 
for higher redshift ($3\times10^{-3}$ Gyr$^{-1}$ at $z=3$) 
and for lower mass ratios ($4\times10^{-3}$ Gyr$^{-1}$ for $M:m\le1:10$).
Thus, the required merger rate at $z>2$ as directly derived from the observed 
data to justify the data themselves in the hypothesis of accretion through
minor mergers, is three orders of magnitude higher than 
the one predicted by models. 
Moreover, such high merger rates are excluded also from the merger rate
derived from the observations of galaxy pairs at high-z.
For instance, L\'opez-Sanjuan et al. (2011) estimate a minor merger rate
R$_{mm}\sim0.04$ Gyr$^{-1}$ at $z\sim0.8$ and they find that it should decrease
with increasing redshift.
Thus, at $z\sim1.5$ we should expect a minor merger rate R$_{mm}<0.04$ 
Gyr$^{-1}$, more than two orders of magnitude lower than the one required 
above.

Some recent studies which find almost exclusively compact ETGs at high-z 
contrary to our observational evidence, consider the possibility that
this difference is due to the higher redshift range covered by their samples.
For instance, Cassata et al. (2010)  and Szomoru et al. (2011) 
estimate the size  of ETGs at $1.5<z<2.5$ using WFC3 data 
(0.13 arcsec/pix vs  FWHM$\sim0.15$ arcsec) finding almost all compact ETGs.
They select (and define) ETGs according to their colors and/or specific star 
formation rate.
A possible suggested explanation for the higher
fraction of normal ETGs observed at lower redshift ($z\sim1-1.5$) 
is that some of the compact ETGs at high-z increase their size between 
$z\sim2.5$ and $z\sim1.5$, i.e in about 1.5 Gyr.
However, if this was the case, following the same reasoning as above, a galaxy 
should have experienced  from five to ten mergers in about 1.5 Gyr, requiring 
a merger rate of 3-7 Gyr$^{-1}$ at $1.5<z<2.5$, even higher than the one 
excluded before.  
It seems then that the different fraction of compact galaxies they 
found with respect to our sample cannot be accounted for by this 
hypothesis but rather it should be attributed to other factors.
For these works,  particular attention should be paid to the different 
criteria used to define and hence to select early-type galaxies.
In this regards, it is worthwhile to mention the work by van der Wel et al.
(2011).
They investigated the stellar structure of massive, quiescent galaxies 
in the redshift range $1.5<z<2.5$ using HST-WFC3 images. 
They found that their effective radii were actually smaller than ETGs in the 
present-day universe with equal stellar mass.
However, they estimate that 65$\pm15$ per cent of the population of massive
quiescent galaxies at $z\sim2$ are disk dominated.
Thus, the hypothesis of an inside-out growth driven by dry merging does not 
seem a viable mechanism to justify the large spread of the effective 
radius and of the stellar mass densities characterizing the population 
of ETGs at $1<z<2$.
To reinforce this conclusion it is also the detection of 
color gradients high-z ETGs (Gargiulo et al. 2011; Guo et al. 2011)
In particular, Gargiulo et al. (2012) detect color gradients in some\
but not all the ETGs of the sample at $1<z<2$.
Most importantly, they find that the presence or the absence of color 
gradients is not related to the compactness of the galaxy: color gradients 
are presents both in compact and in normal ETGs as well as some of the 
ETGs, both compact and normal, do not show color gradients. 
Necessarily, inside-out growth driven by dry merging cannot be assumed as
a general scheme of accretion but a different scenario must be hypothesized.
We further analyzed the properties of high-z ETGs in order to gain new
insights and constraints on the possible mechanism(s) able to 
produce the observed properties.

\section{Central stellar mass density and scaling relations}  
\begin{figure*}
\begin{center}
%\hskip -0.2truecm
\includegraphics[width=8.5cm]{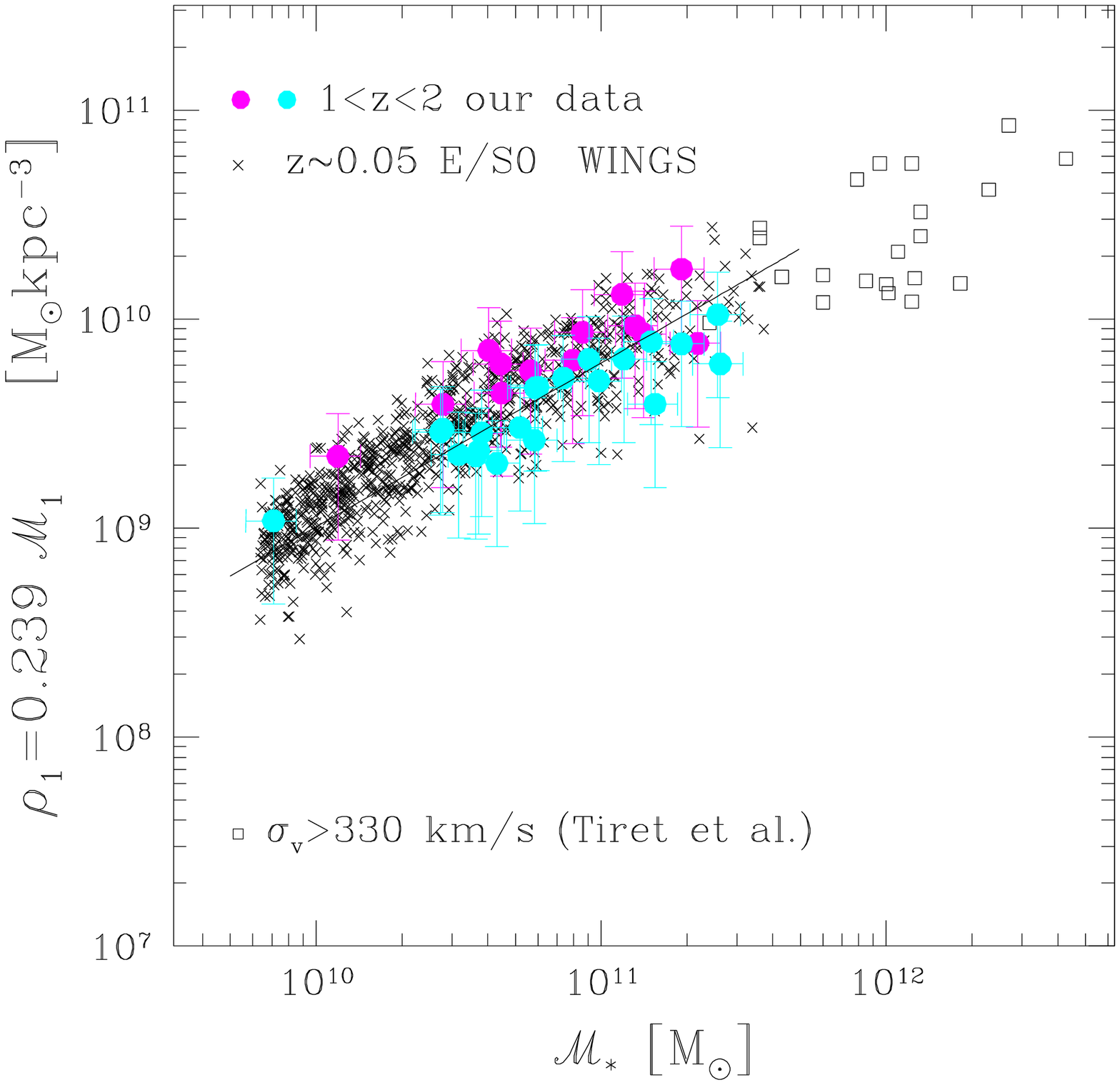} 
\includegraphics[width=8.5cm]{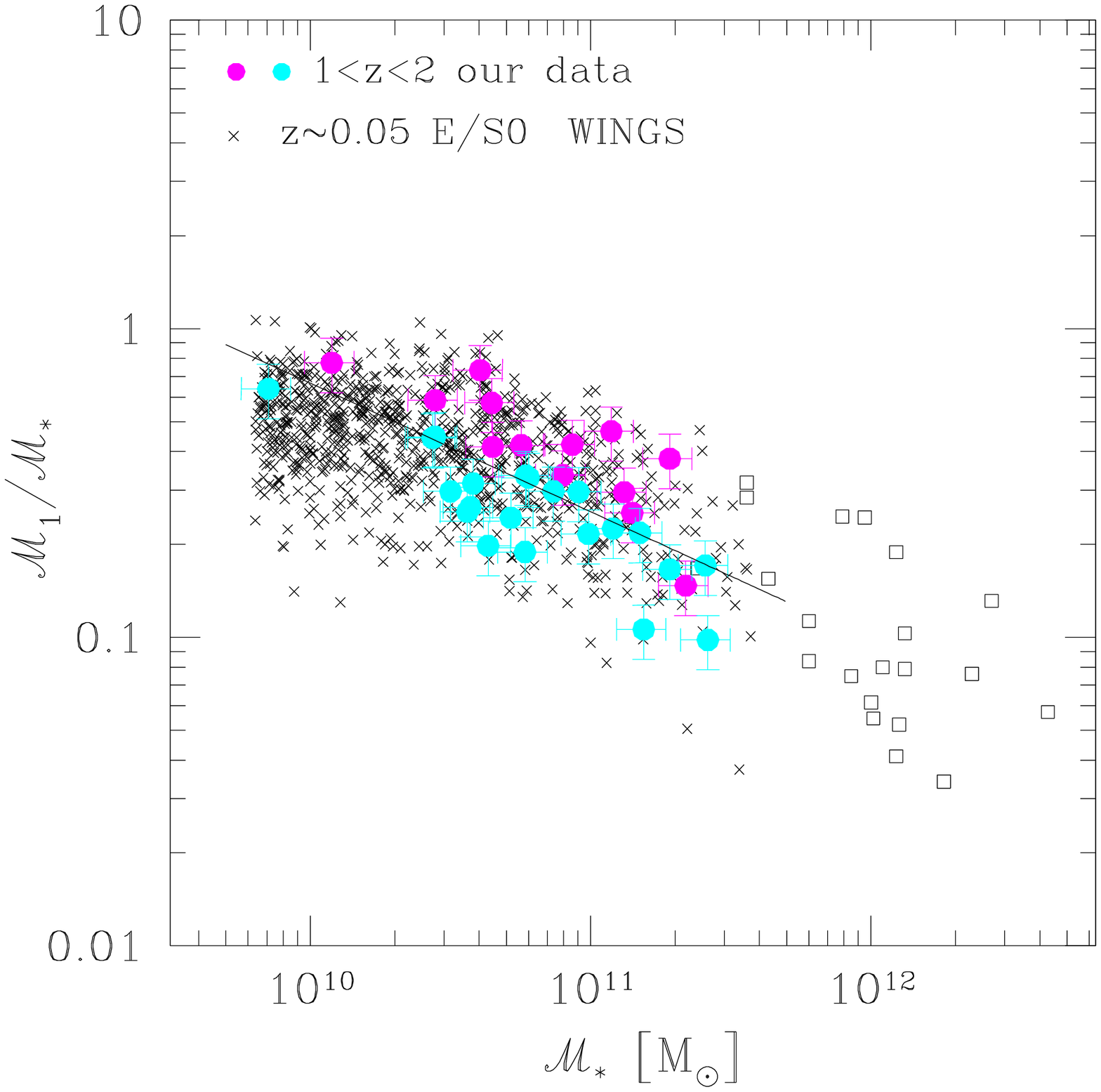} 
\vskip -0.5truecm
\caption{Left - The central stellar mass density $\rho_1$ is plotted versus
the total stellar mass $\mathcal{M}_*$ for our sample of high-z ETGs (large
filled magenta and cyan circles) and for the WINGS sample of local ETGs
(black crosses).
These two quantities are tightly correlated independently of the redshift and the
compactness of the galaxies.
The correlation is well represented by eq. (4).
Right - The fraction of the stellar mass within 1 kpc radius is plotted versus
the total stellar mass of the galaxy.
Symbols are the same as in the left panel.
The existence of the correlation shown in the left panel and of the form of eq. (4)
implies the correlation shown in this plot: the fraction of stellar mass contained
within 1 kpc radius decreases with increasing total stellar mass of the system
(eq. (5)).}
\end{center}
\end{figure*}

In the left  panel of Fig. 7 the central stellar mass density $\rho_1$
is plotted versus the stellar mass $\mathcal{M}_*$ of the galaxies.
It is evident the tight correlation between the two quantities.
The correlation is well reproduced by the relation
\begin{equation}
\rho_1=2.35(\pm0.3)\times 10^3 \mathcal{M}_*^{0.58\pm0.16)}.
\end{equation}
for our high-z ETGs (filled circles).
We note that the local sample of ETGs (crosses) follow the same correlation
with a best-fitting relation possibly slightly steeper 
($\rho_1\propto\mathcal{M}_*^{0.73}$).
It is worth noting that Tiret et al. (2011) found an opposite correlation
using their data set ($\rho_1\propto\mathcal{M}_*^{-0.25}$), in contrast 
with our result and also with what the data of the other authors they used  
show (see their Fig. 6). 
There are two main differences between their work and the others which
could justify this discrepancy.
The first one is that they consider the sample of super-massive ellipticals
($0.3-5\times10^{12}$ M$_\odot$) with velocity 
dispersion larger than 330 km s$^{-1}$  (Bernardi et al. 2006), while the 
other samples (as well as our own) include galaxies less massive with no 
cuts in velocity dispersion.
The other difference is the stellar mass estimator used to derive the density 
within 1 kpc.
Indeed, while we and the other authors have used the light profile to
derive the fraction of stellar mass within a fixed radius, they
use a dynamical mass computed by solving the Jeans equation.
To directly verify whether the different stellar mass range 
characterizing their sample can justify the different result, we derived
the central stellar mass density of their ETGs according to our method 
(using eq. 2) using the stellar mass they obtain from stellar population 
model (column 6 in their Tab. 2).
Their sample is represented in figures 7 by open squares.
It is evident in this case the agreement of their data with the others 
data in all the 
cases, showing that the different mass range and the selection 
in velocity dispersion are not the reasons of their opposite correlation.
We believe that the use of the dynamical mass they define is the 
reason of the different result. 
{ 
Actually, their Fig. 4 shows the large scatter which exists between the 
dynamical mass and the stellar mass they derived and a relation between
them slightly steeper than 1:1 which could be the reasons of the different 
behavior they find.
}
The result obtained from this comparison shows that the correlation 
between the central stellar mass density and the stellar mass 
discussed above and the resulting considerations are valid for the 
population of ETGs independently of redshift, of mass and of velocity 
dispersion ranges considered.

Because $\rho_1=0.239\mathcal{M}_1$, it follows that also the central stellar mass  
increases with the total stellar mass of the galaxy according to 
the same scaling relation, $\mathcal{M}_1\propto\mathcal{M}_*^{0.58}$.
Furthermore, from the above relation it follows that the fraction of the 
central
stellar mass decreases with mass as 
\begin{equation}
\mathcal{M}_1/\mathcal{M}_*=9.85\mathcal{M}_*^{-0.42}
\end{equation}
Hence, if the stellar mass of the galaxy increases by a factor two then the stellar
mass and the density within 1 kpc increase more slowly, by a factor 1.5.
Consequently, the fraction of stellar mass contained within 1 kpc, 
$\mathcal{M}_1/\mathcal{M}_*$, decreases by a factor 1.3 when 
the total mass double as shown in the right panel of Fig. 7.
Finally, it is interesting to note that since the effective radius of
ETGs increases with mass as R$_e\propto\mathcal{M}_*^{\sim0.56}$ 
(e.g. Shen et al. 2003) and $\rho_1\propto\mathcal{M}_*^{\sim0.6}$ (eq. 4)
it follows that $\rho_1$ is nearly independent of R$_e$, as it is.
This is another and independent way to show that $\rho_1$ and $\rho_e$
are nearly uncorrelated, independent of redshift and of assembly
history since the relations used to derive this correlation are valid 
both for high-z ETGs and for local ETGs independently of their compactness.
Moreover, this indirectly shows that the small variation of the central
stellar mass density of ETGs with respect to the effective mass density 
is an intrinsic characteristic of the Sersic's profile
and has no regards with the inside-out growth.

In all the cases considered above there are no differences between the 
high-z sample and the local sample in spite of the extra 9 Gyrs for
local ETGs compared with high-z ETGs.
All the ETGs follow the tight correlations shown above independently of 
their compactness and of their redshift.
Hence, whatever evolution experienced by ETGs during the last 9-10 Gyr of 
their life, must be able to leave their properties unchanged.
Such properties, already in place at $z\sim1.5-2$, must necessarily  
originate at earlier epoch, during the first 3 Gyr.
The large spread of the effective stellar mass density equally
observed in the ETGs at $z\sim1.5$ and in the local ETGs representing the
large variation of R$_e$ at fixed stellar mass, must  
originate in the early-phases of their formation.
And the mechanism(s) responsible for mass growth of ETGs must proceed 
according to the above scaling relations.

\begin{table*}
\caption{
For each galaxy of the sample we report the spectroscopic redshift, the total apparent magnitude 
F850LP$_{fit}$
resulting from the best fitting profile on the HST-ACS images, the effective radius R$_e$, the S\'ersic 
index $n$, 
the effective stellar mass density $\rho_e$ measured within the effective radius, 
the central stellar mass density $\rho_1$ measured within 1 kpc radius, the total stellar mass  
$\mathcal{M}_*$ and the stellar mass $\mathcal{M}_1$ within 1 kpc, the mean age of the stellar population.
We remind that $\rho_1=0.239\mathcal{M}_1$. { The last column (OII)$_{em}$ reports those galaxies
showing OII emission.}
}
\centerline{
\begin{tabular}{rrrcrrrrrrr}
\hline
\hline
 ID & $z_{spec}$ &F850LP$_{fit}$& R$_e$ & n & log$\rho_e$& log$\rho_1$
 &log$\mathcal{M}_*$&log$\mathcal{M}_1$& Age& (OII)$_{em}$\\
    &            & [mag]        &[kpc]  &   &[M$_{\odot}$ kpc$^{-3}$]&[M$_{\odot}$ kpc$^{-3}$]& 
    [M$_{\odot}$] & [M$_{\odot}$]& Gyr& \\
  \hline                                                         
   472& 1.921&  23.84&   0.51$\pm$0.03&   2.4&   10.51&   9.80&   10.56&  10.42&   1.02& -- \\ 
  2361& 1.609&  23.31&   1.14$\pm$0.10&   4.0&    9.90&  10.03&   10.99&  10.66&   3.25& (1) \\ 
  2148& 1.609&  22.75&   1.64$\pm$0.15&   4.8&    9.61&  10.13&   11.17&  10.75&   3.50& -- \\ 
  2111& 1.610&  23.33&   0.78$\pm$0.04&   3.5&    9.97&   9.71&   10.57&  10.33&   1.02& -- \\ 
    12& 1.123&  21.08&   1.38$\pm$0.03&   4.9&    9.47&   9.81&   10.81&  10.43&   2.40& -- \\ 
 11322& 1.032&  22.08&   0.41$\pm$0.01&   3.0&   10.25&   9.27&   10.00&   9.89&   1.02& -- \\ 
  2543& 1.612&  23.91&   2.06$\pm$0.32&   3.2&    9.13&   9.84&   11.00&  10.47&   3.00& (1) \\ 
  2355& 1.610&  23.67&   0.80$\pm$0.03&   2.1&    9.78&   9.55&   10.40&  10.17&   0.90& -- \\ 
     3& 1.044&  20.98&   1.39$\pm$0.02&   4.9&    9.26&   9.62&   10.62&  10.24&   2.00& -- \\ 
 12294& 1.215&  21.55&   1.24$\pm$0.01&   1.8&    9.40&   9.60&   10.60&  10.22&   1.02& (2) \\ 
  2196& 1.614&  22.92&   1.99$\pm$0.24&   4.8&    8.93&   9.65&   10.75&  10.27&   1.68& -- \\ 
     4& 0.964&  19.98&   3.63$\pm$0.03&   2.4&    8.71&   9.86&   11.31&  10.48&   3.00& -- \\ 
 11804& 1.910&  22.97&   2.87$\pm$0.16&   4.5&    8.71&   9.78&   11.00&  10.41&   0.90& -- \\ 
 11539& 1.096&  20.20&   4.25$\pm$0.44&   4.0&    8.55&   9.97&   11.36&  10.60&   2.75& -- \\ 
    11& 1.096&  21.46&   2.42$\pm$0.25&   5.0&    8.88&   9.81&   10.96&  10.43&   3.25& (3) \\ 
  2286& 1.604&  23.72&   1.17$\pm$0.08&   2.2&    9.25&   9.40&   10.38&  10.02&   1.14& (1) \\ 
  9369& 1.297&  22.27&   1.98$\pm$0.15&   4.7&    8.92&   9.64&   10.74&  10.26&   2.30& -- \\ 
 12327& 1.097&  22.15&   1.24$\pm$0.06&   4.8&    9.13&   9.36&   10.33&   9.98&   1.90& -- \\ 
     8& 1.125&  21.48&   2.07$\pm$0.04&   5.0&    8.79&   9.55&   10.66&  10.18&   2.10& -- \\ 
 11888& 1.039&  20.97&   2.37$\pm$0.05&   4.8&    8.69&   9.59&   10.74&  10.21&   2.50& -- \\ 
    20& 1.022&  20.41&   3.85$\pm$0.09&   5.4&    8.32&   9.72&   11.00&  10.34&   2.75& -- \\ 
 12965& 1.019&  22.72&   0.61$\pm$0.02&   3.8&    9.49&   8.95&    9.77&   9.58&   1.14& -- \\ 
     1& 1.089&  20.34&   3.23$\pm$0.05&   3.9&    8.41&   9.57&   10.86&  10.19&   0.81& (3) \\ 
    13& 0.980&  19.83&   3.75$\pm$0.04&   5.5&    8.30&   9.67&   10.94&  10.29&   1.61& -- \\ 
     2& 0.964&  19.69&   5.69$\pm$0.13&   5.7&    7.92&   9.71&   11.11&  10.33&   2.60& -- \\ 
  2239& 1.415&  23.26&   2.06$\pm$0.14&   2.0&    8.62&   9.27&   10.48&   9.89&   1.14& (1) \\ 
 12789& 1.221&  22.55&   2.21$\pm$0.22&   5.0&    8.47&   9.31&   10.43&   9.93&   2.40& -- \\ 
    23& 1.041&  21.19&   2.80$\pm$0.08&   3.9&    8.29&   9.32&   10.56&   9.94&   1.90& -- \\ 
  9066& 1.188&  22.45&   2.43$\pm$0.16&   3.5&    8.36&   9.24&   10.44&   9.86&   2.50& -- \\ 
 12000& 1.222&  22.06&   2.41$\pm$0.08&   5.0&    8.35&   9.27&   10.42&   9.89&   1.14& (2)\\ 
     7& 1.135&  20.28&   9.29$\pm$0.93&   5.0&    7.49&   9.69&   11.32&  10.31&   2.60& -- \\ 
 10414& 1.170&  21.87&   3.47$\pm$0.22&   3.7&    7.91&   9.13&   10.46&   9.75&   2.40& -- \\ 
 12434& 1.330&  21.31&   7.88$\pm$0.74&   4.6&    7.32&   9.34&   10.93&   9.96&   1.80& -- \\ 
    14& 0.984&  20.73&   4.40$\pm$0.44&   5.0&    7.69&   9.20&   10.54&   9.82&   1.70& -- \\ 
\hline									 
\hline									 
\end{tabular}								 
}	
(1) Cimatti et al. (2008); (2) di Serego Alighieri et al. (2005); (3) van der Wel et al. (2005)								 
\end{table*}

\section{Summary and conclusions}  
We studied the relations between the stellar mass density within the effective
radius $\rho_e$ and the stellar mass density within a fixed radius of
1 kpc  $\rho_1$ for a complete sample
of 34 ETGs at $0.9<z_{spec}<2$ with stellar masses in the range 
$0.6\times10^{10}-3\times10^{11}$ M$_\odot$.
As previously found by other authors (e.g. Tiret et al. 2011; Bezanson et al.
2009) we find that the central stellar mass density $\rho_1$ of ETGs 
varies just by a factor of two and it is similar to the one
of local ETGs.
This property of the high-z ETGs was supposed to be the sign of the
spheroid formation scenario in which ETGs assemble at high-z as 
compact/dense spheroids and then add a low stellar mass density envelope
growing efficiently in size and less in mass.
The central stellar mass density and the stellar mass of the galaxy 
would remain nearly constant while the 
effective stellar mass density would strongly decrease across the time
as the effective radius increases.
However, we find that the effective stellar mass density $\rho_e$ of 
high-z ETGs varies by almost 3 orders of magnitude exactly as the 
local ETGs.
We find indeed that high-z and low-z ETGs selected in the same stellar
mass range populate the same locus in the [$\rho_e$;$\rho_1$] plane
and that the distribution of the effective stellar mass density 
did not change since $z\sim1.5$, i.e. in the last 9-10 Gyr.
This implies that the large spread of $\rho_e$ observed both in the
local universe and up to $z\sim1.5-2$ must originate earlier, at $z>2$.
Moreover, we show that the small variation of the central stellar mass density
with respect to the large variation of the effective density does not depend 
on redshift and/or on the past assembly history.
On the contrary, this feature is characteristic of the Sersic profile.
Thus, the relation between the stellar mass densities of ETGs is not 
representative and cannot be used to probe their past 
assembly history.
It has no connection with the inside-out growth.
We investigated whether the spread of the effective radius and of the
effective stellar mass density observed in ETGs up to $\sim1.5$ 
can be the sign of the inside-out growth taken place at higher redshift
($z>2$). 
We find that the number of mergers needed to justify at $z\sim1.5$ a spread
of a factor 4 in R$_e$ or about 60 in $\rho_e$ is from five to ten in the 
redshift range $2<z<6$, corresponding to a merger rate of about 2-4 Gyr$^{-1}$.
This merger rate is three orders of magnitude higher than the one predicted
by models and than the one derived from the observations of high-z 
galaxy pairs.

{ As also discussed in Saracco et al. (2011), a scenario where the
main starburst triggered by the dissipative gas-rich merging is the main 
driver of the stellar mass growth, could qualitatively account for the 
observations when supposing that gas cooling can modulate the star formation.
Indeed, assuming that dissipative gas-rich merger at high-z ($z>4-5$) is the 
main mechanism of spheroids formation, compact ETGs would be a natural 
consequence of this mechanism when most of the gas at disposal is burned 
in the central starburst, that is when the gas is  sufficiently cold to 
collapse toward the center and to ignite the main starburst.
On the contrary, when the gas of the progenitors is not sufficiently cold and 
homogeneous the resulting starburst would not be short, central and intense 
but longer and possibly composed of many episodes.
The cooling time of the gas clouds and their orbital parameters
could modulate the rate at which the starburst(s) are ignited, stoked
and distributed across the galaxy.
}

We find and define a tight correlation between the central stellar mass 
density of ETGs and the total stellar mass in the sense that the central 
stellar mass density increases with mass as $\mathcal{M}_*^{\sim0.6}$.
We find that this correlation is general and it is valid for the whole 
population of ETGs, independently of  redshift, mass and/or of velocity 
dispersion ranges considered.
Given this correlation it follows also that the fraction of the central
stellar mass of ETGs decreases with mass. 
 
The large spread of the effective stellar mass density equally
observed in the ETGs at $z\sim1.5$ and in the local ETGs representing the
large variation of R$_e$ at fixed stellar mass, must obviously 
originates at $z>2$.
The fact that this property apparently did not change in the subsequent
9-10 Gyr as it is equally observed in the local ETGs, suggest
that it must have originated during the early phases of their formation.
Moreover, whatever the evolution ETGs underwent during the last 9-10 Gyr of 
their life, such evolution must be able to leave unchanged their properties 
and the correlation found.

\section*{Acknowledgments}
This work is based on observations made with the ESO telescopes at the Paranal 
Observatory and with the NASA/ESA Hubble Space Telescope, obtained from the 
data archive at the Space Telescope Science Institute which is operated by 
the Association of Universities for Research in Astronomy. 
We thank B. Poggianti, A. Moretti, T. Valentinuzzi and the WINGS team for
having provided us with their data. 
We would like to thank the referee, John Stott, for useful and
constructive comments which improved the quality of the paper.
This work has received financial support from ASI-INAF (contract I/009/010/0). 

\section{References}

\noindent Bernardi, M., et al. 2006, AJ, 131, 2018

\noindent Bernardi, M., Hyde, J. B., Fritz, A., Sheth, R. K., Gebhardt, K., 
Nichol, R. C. A., 2008, MNRAS, 391, 1191

\noindent Bezanson R., van Dokkum P. G., Tal T., Marchesini D., Kriek M.,

\noindent Buitrago F., Trujillo I., Conselice C. J., Bouwens R. J., Dickinson
M., Yan H. 2008, ApJ, 687, L61

\noindent Cameron E., Carollo C. M., Oesch P. A., Bouwens R. J., Illingworth
G. D., Trenti M., Labb\'e I., Magee D., 2011, ApJ, 743, 146

\noindent Capaccioli, M. 1989, in World of Galaxies (Le Monde des Galaxies),
 208

\noindent Cassata P., et al. 2010, ApJ, 714, L79

\noindent Cassata P., et al. 2011, ApJ, 743, 96

\noindent Cimatti A., et al. 2008, A\&A, 482, 21

\noindent Ciotti L., 1991, A\&A, 249, 99

\noindent Collins  C. A., et al. 2009, Nature 458, 603

\noindent Daddi E., Renzini A., Pirzkal N., et al. 2005, ApJ, 626, 680

\noindent Damjanov I., McCarthy P. J., Abraham R. G., et al. 2009, ApJ, 
695, 101

\noindent Damjanov I., et al. 2011, ApJ, 739, L44

\noindent De Lucia G., Springel V., White S. D. M., Croton D., 
Kauffmann G. 2006, MNRAS, 366, 499

\noindent di Serego Alighieri S., et al. 2005, A\&A, 442, 125

\noindent Fasano G., et al. 2006, A\&A, 445, 805

\noindent Fasano G., et al. 2012, MNRAS 420, 926

\noindent Gargiulo A., Saracco P., Longhetti M. 2011, MNRAS, 412, 1804

\noindent Gargiulo A., Saracco P., Longhetti M. La Barbera F. 2012, MNRAS, 
submitted

\noindent Giavalisco M., et al. 2004, ApJ 600, L103

\noindent Graham A. W., Driver S. P., 2005, PASA, 22, 118

\noindent Guo Y., et al. 2011, ApJ, 735, 18

\noindent Hopkins, P., Bundy, K., Murray N., Quataert E., Lauer T. R.,
Ma C.-P. 2009, MNRAS, 398, 898

\noindent Hopkins, P., et al. 2010, ApJ, 715, 202

\noindent Kennicutt R. C. Jr., 1998, ARA\&A 36, 189

\noindent Khochfar S., Burkert A., 2003, ApJ, 597, L11

\noindent Khochfar S., Silk J. 2006a, ApJ, 648, L21 

\noindent Longhetti M., Saracco P., Severgnini P., et al., 2007, MNRAS, 374, 614

\noindent L\'opez-Sanjuan C., et al. 2011, A\&A, 530, 20

\noindent Mancini, C., Daddi E., Renzini A., et al. 2010, MNRAS, 401, 933

\noindent McGrath E., Stockton A., Canalizo G., Iye M., Maihara T. 2008,
ApJ, 682, 303

\noindent Naab T., Johansson P. H., Ostriker J. P., Efstathiou G. 2007, 
ApJ, 658, 710

\noindent Naab T., Johansson P. H., Ostriker J. P. 2009, ApJ, 699, L178

\noindent Newman A. B., Ellis R. S., Treu T., Bundy K. 2010, ApJ, 717, L103

\noindent Nair P., van den Bergh S., Abraham R., 2011, ApJ, 734, L31

\noindent Nipoti C., Treu T., Auger M. W., Bolton A. S. 2009, ApJ, 706, L86

\noindent Peng C. Y., Ho L. C., Impey C. D., Rix H.-W., 2002, AJ 124, 266

\noindent Pignatelli E., Fasano G., Dassata P., 2006, A\&A, 446, 373

\noindent Pozzetti L., et al. 2010, A\&A 523, 13 

\noindent Renzini A., 2006, ARA\&A, 44, 141

\noindent Saracco P., Longhetti M., Andreon S., 2009, MNRAS, 392, 718

\noindent Saracco P., Longhetti M., Gargiulo A. 2010, MNRAS, 408, L21

\noindent Saracco P., Longhetti M., Gargiulo A. 2011, MNRAS, 412, 2707

\noindent Shen S., et al., 2003, MNRAS, 343, 978

\noindent Sommer-Larsen J., Toft S., 2010, ApJ, 721, 1755

\noindent Stott J. P., et al. 2010, ApJ 718, 23

\noindent Stott J. P., Collins C. A., Burke C., Hamilton-Morris V., Smith G.
P., 2011, MNRAS, 414, 445

\noindent Szomoru D., Franx, M., van Dokkum P. G., 2011, ApJ 735, L22

\noindent Thomas D., Maraston C., Bender R., Mendes de Oliveira C. 2005,
ApJ, 621, 673

\noindent Tiret O., Salucci P., Bernardi M., Maraston C., Pforr J. 2011,
MNRAS, 411, 1435

\noindent Trujillo I., Feulner G., Goranova Y., et al. 2006, MNRAS, 373, L36

\noindent Trujillo I., Conselice C. J., Bundy K., Cooper M. C., Eisenhardt P,
Ellis R. S., 2007, MNRAS, 382, 109

\noindent Trujillo I., Ferreras I., de La Rosa I. G. 2011, MNRAS, 415, 3903

\noindent Valentinuzzi P., et al. 2010a, ApJ, 712, 226

\noindent Valentinuzzi P., et al. 2010b, ApJ, 721, L19

\noindent van der Wel A., M. Franx, P. G. van Dokkum, H.-W. Rix, G. D.
Illingworth, P. Rosati, 2005, ApJ, 631, 145

\noindent van der Wel A., Holden B. P., Zirm A. W., Franx, M., Rettura, A.,
Illingworth G. D., Ford H. C., 2008, ApJ, 688, 48

\noindent van der Wel A., Bell E. F., van den Bosch F. C., Gallazzi A., Rix
H. 2009, ApJ, 698, 1232

\noindent van der Wel A., et al. 2011, ApJ 730, 38

\noindent van de Sande J., et al. 2011, ApJ 736, L9

\noindent van Dokkum P. G., et al. 2008, ApJ, 677, L5

\noindent van Dokkum P. G., et al. 2010, ApJ, 709, 1018

\noindent Wuyts S., et al. 2010, ApJ 722, 1666

\label{lastpage}
\end{document}